\documentclass[submission,copyright,creativecommons]{eptcs}
\providecommand{\event}{SOS 2007} 
\usepackage{breakurl}             
\usepackage{amsmath,amssymb,amsthm}

\newtheorem{theorem}{Theorem}

\newcommand{\mb}{\mathbb}

\title{The Problem of Analogical Inference in Inductive Logic}
\author{Simon M. Huttegger
\institute{Department of Logic and Philosophy of Science}
\institute{School of Social Sciences\\
University of California, Irvine\\
Irvine, CA-92617, USA}
\email{shuttegg@uci.edu}
}
\def\titlerunning{Analogical Inference in Inductive Logic}
\def\authorrunning{S. M. Huttegger}
\begin{document}
\maketitle

\begin{abstract}
We consider one problem that was largely left open by Rudolf Carnap in his work on inductive logic, the problem of analogical inference. After discussing some previous attempts to solve this problem, we propose a new solution that is based on the ideas of Bruno de Finetti on probabilistic symmetries. We explain how our new inductive logic can be developed within the Carnapian paradigm of inductive logic---deriving an inductive rule from a set of simple postulates about the observational process---and discuss some of its properties.
\end{abstract}

\section{Introduction}

The logical empiricist movement is often associated with using deductive logic to understand scientific reasoning. But Rudolf Carnap actually favored an inductive approach, starting with his work on inductive logic in the 1940s. Carnapian inductive logic can be thought of as a branch of probability theory that is especially concerned with predictive probabilities---the probability of future observations given past observations. Carnap spent much of the last thirty years of his life on developing an inductive logic, but even in his posthumously published works he considered the subject to be wide open to further investigations. The open problem that I wish to consider in this paper is the problem of analogical inference, which hasn't received a satisfactory answer in Carnap's original system. I shall review some of the attempts to develop an analogical inductive logic in \S \ref{analogy}. In order to set the stage, I briefly describe Carnap's program in \S \ref{carnap} and point to its connections with de Finetti's theory of inductive inference in \S \ref{finetti}. Considering de Finetti is particularly important since he provides an alternative route to analogical inference. In \S \ref{partial} I discuss an especially interesting probabilistic symmetry that allows for a certain form of analogical inference. Finally, in \S \ref{logic} I introduce a new analogical inductive logic based on that symmetry and discuss some of its properties.

\section{Carnap's Program} \label{carnap}

Carnap's program for developing an inductive logic as described in his `Logical Foundations of Probability' \cite{Carnap1950} was brought to a tentative conclusion in the posthumously published `A Basic System of Inductive Logic' \cite{Carnap1971b,Carnap1980}. Carnapian inductive logic aims at finding rational foundations for the kind of inductive inferences that are used in scientific investigations. The classic example of such an inference in the tradition of Bayes and Laplace is the predictive probability of events, such as future coin flips based on past observations of coin flips. Carnap viewed all inductive inference problems as being essentially reducible to this type of inference \cite{Carnap1950}.\footnote{See \cite{Zabell2011} for an excellent overview for the development of Carnapian inductive logic.} 

Of particular importance for Carnap are predictive probabilities based on the relative frequencies of events. For example, after observing a number of throws of a die, the predictive probability of observing a six with the next throw usually is judged to be approximately equal to the relative frequency of sixes. In his systems of inductive logic, Carnap tries to explicate the foundations of this simplest kind of inductive inference.

Independently of Carnap's program, a similar approach was developed more than two decades earlier by the Cambridge logician W. E. Johnson \cite{Johnson1924,Johnson1932}. Johnson's main contribution was only published posthumously and contained a number of gaps, which were closed by Sandy Zabell \cite{Zabell1982}, who also generalized Johnson's approach to a theory that is essentially equivalent to Carnap's basic system of inductive logic. I'm going to follow Zabell's elegant treatment because it ties in neatly with the work of Bruno de Finetti (see the next section).\footnote{Kuipers \cite{Kuipers1978} gives an overview of the mathematical aspects of Carnap's theory.} 

The basic postulate in this theory of inductive inference is a symmetry requirement known as `exchangeability' (called the `permutation postulate' by Johnson). Suppose that there is a finite sequence of random variables $X_1, \hdots, X_n$ representing observations (e.g. coin flips), and let their probability law be $\mb P$. Like Carnap, we assume that the random variables can take on only a finite number of values. Then $\mb P$ is {\it exchangeable} if it is invariant under permutations of outcomes; that is, 
$$
\mb P[X_1 = x_1, \hdots, X_n = x_n] = \mb P[X_1 = x_{\sigma(1)}, \hdots, X_n = x_{\sigma(n)}]
$$
for every permutation $\sigma$ of $\{1, \hdots, n\}$. This allows us to define exchangeable probabilities of infinite sequences $X_1, X_2, \hdots$ as those for which every finite initial sequence is exchangeable. For simplicity,  the sequence of random variables is often called exchangeable without referring to its probability law.

Both Johnson and Carnap use a requirement for predictive probabilities that is often called `Johnson's sufficientness postulate'. This postulate says that predictive probabilities for $i$ basically only depend on the past relative frequency of $i$; i.e., there is a function $f$ such that
\begin{equation} \label{johnson}
\mb P[X_{n+1} = i | X_1, \hdots, X_n] = f_i(n_i,n).
\end{equation}
Johnson's sufficientness postulate judges information about types other than $i$ to be irrelevant for the predictive probability of $i$---a point that is going to be important for the problem of analogical inference.

Finally, in order for conditional probabilities to be well defined, a regularity postulate is assumed to the effect that each finite initial sequence of outcomes has positive probability. It is then possible to show that the predictive probability of any outcome is equal to its relative frequency modulo some prior parameters. More specifically, if trials are not independent, then there exist parameters $\alpha_j$ for each outcome $j$ such that for all $n$ and $i$
\begin{equation} \label{basic}
\mb P[X_{n+1} = i | X_1, \hdots, X_n] = \frac{n_i + \alpha_i}{n + \sum_j \alpha_j}.
\end{equation}
(If trials are independent, then there is no learning from experience.) Here $n_i$ is the number of times outcome $i$ is observed in the first $n$ trials. The parameters $\alpha_j$ are either all positive or all negative; they must be positive if the sequence of observations is infinite exchangeable (see \cite{Zabell1982} for a thorough discussion). The rule given by \eqref{basic} is called a `generalized rule of succession' (after Laplace's special `rule of succession'). A generalized rule of succession expresses a mode of learning from experience. Experiences are given by past observations of outcomes, and past observations lead to predictive probabilities for future outcomes.

The inductive logic given in \eqref{basic} is equivalent to Carnap's mature basic system of inductive logic, also known as the `$\lambda-\gamma$-continuum of inductive methods'. The system championed in his 1950 book is much more restricted \cite{Carnap1950}. It requires that all $\alpha_j = 1$, meaning that all outcomes are judged to be equally probable prior to any observations. In his later `A Continuum of Inductive Methods' \cite{Carnap1952}, Carnap generalized this restricted system to one with a weight $\lambda$ which regulates the effect of the equally probable prior weights. The basic system \eqref{basic} extends this to arbitrary prior weights.

Especially in his early work on inductive logic, Carnap thought of symmetry principles such as exchangeability as requirements of rationality. The idea---familiar from justifications for Laplace's principle of indifference---is that certain probabilistic symmetries should hold whenever one does not have any knowledge about the relevant underlying structure. For instance, in the absence of any evidence concerning the order of outcomes you should assume exchangeability. We will see that interpreting symmetry principles in this way puts significant constraints on how to include analogy effects into inductive logic, while the approach discussed in the next section allows for a greater variety of inductive logics.

\section{De Finetti's Program} \label{finetti}

Bruno de Finetti is famous for his foundational work on probability theory and inductive inference. The latter is of special importance to us here. The most fundamental result in this arena is de Finetti's representation theorem for  exchangeable sequences \cite{Finetti1937}. Exchangeability is important because it captures one of the classic situations of statistics---i.i.d. trials with unknown parameters. This is what is shown by de Finetti's representation theorem. Suppose, for example, that $X_i$ records whether the $i$th toss of a coin flip came up heads or tails, and that the infinite sequence $X_1, X_2, \hdots$ is exchangeable.  de Finetti proved that this is equivalent to the probability of finite sequences of heads and tails being a mixture of i.i.d. binomial trials with unknown bias of the coin:\footnote{For finite forms of this result, see \cite{Diaconis1980}.}
\begin{equation} \label{rep}
\mb P[X_1 = x_1, \hdots, X_n = x_n] = \int_0^1 p^h (1-p)^{n-h} d \mu(p)
\end{equation}
 (Here, $p$ is the bias for heads, $\mu$ is a uniquely determined prior over biases and $h$ is the number of heads in the first $n$ trials.) This theorem has profound consequences for the philosophy of probability and for inductive inference \cite{Zabell1989}.  Specifically, if the prior $\mu$ in the representation is a Beta distribution (or, in the more gneral case of finitely many types of outcomes, a Dirichlet distribution), then
$$
\mb P[X_{n+1} = i | X_1, \hdots, X_n] = \frac{n_i + \alpha_i}{n+ \sum_j \alpha_j},
$$  
where $\alpha_i,\alpha_j$ are nonnegative parameters determining the Dirichlet distribution.  This is equivalent to the Carnapian inductive logic given in \eqref{basic}. One difference between the two approaches lies in the underlying axiomatic foundations. In de Finetti's case, it is given by (i) the assumption of exchangeability and (ii) the assumption that the mixing prior in the representation $\mu$ is a Dirichlet distribution. In the Johnson-Carnap approach there is no appeal to the de Finetti representation.

There is also an important interpretive issue that separates the early work of Carnap from de Finetti's probabilistic epistemology (in his later work Carnap is closer to de Finetti's views). de Finetti did not view exchangeability or other symmetry requirements as postulates of rationality. According to him, exchangeability is a personal judgement of an epistemic agent as to the basic structure of a learning situation. Such a judgement does not arise from the lack of knowledge but presupposes knowledge about an epistemic situation. 

This view of symmetry assumptions has two important consequences, one epistemological and one formal. In the first place, for de Finetti and his followers the justification of  generalized rules of succession is only a relative one. An agent should make inductive inferences provided that she assumes certain underlying symmetries about the learning situation. This is unlike the objective Bayesian tradition---which includes Bayes, Laplace, Keynes, the early work on inductive logic by Carnap, and others---where symmetry assumptions themselves are viewed not just as assumptions that one may make, but as principles every rational agent has to make under certain conditions. 

de Finetti's probabilistic epistemology is thus distinctly non-foundationalist. There is no bedrock of initial epistemic judgements that would endow all their consequences with full rationality because they are themselves requirements of rationality. For de Finetti, rationality is instead to be found in the interplay of {\it inductive assumptions}, such as Johnson's sufficientness postulate or exchangeability, and rules for learning from observations. If you use such an inductive rule but deny its underlying assumptions, you are simply inconsistent. So, de Finetti requires a kind of relative rationality: learning from experience should be compatible with those inductive assumptions that are judged to be true.

The second consequence of de Finetti's view of symmetry assumptions lifts constraints from inductive logic. If assumptions such as exchangeability are not thought of as requirements of rationality but as personal judgements, then one might consider other kinds of symmetries whenever exchangeability does not seem appropriate. This led de Finetti to study `partial exchangeability' \cite{Finetti1938,Finetti1959,Diaconis1980}. One kind of partial exchangeability, known as `Markov exchangeability', allows outcomes to depend on previous trials \cite{Diaconis1980a,Fortini2002,Freedman1962,Kuipers1988,Skyrms1991,Zabell1995}. The type of partial exchangeability most relevant to our analogical inductive logic was investigated by de Finetti himself \cite{Finetti1938,Finetti1959}. Consider a situation where outcomes can be of different types; e.g., coin flips with two coins, or medical trials with men and women. Then one may not be willing to judge outcomes to be exchangeable across types but only within types. There is a representation theorem for this kind of partial exchangeability, from which predictive conditional probabilities can be derived \cite{Finetti1938}. The representation is very similar to \eqref{rep}. Probabilities are again mixtures of independent trials, but now trials need not be identically distributed; they are identically distributed within types, but need not be so be across types.

de Finetti viewed partial exchangeability as a type of analogical inference. Take the example of flipping two coins. The coins are judged to be similar but not indistinguishable from each other. Because of the analogy between the two coins, observations from one coin should have some influence on predictions for the other coin. The analogy comes from particular prior distributions on the chances in the mixture of the representation theorem. The biases of the two coins may be chosen dependently, but then trials are independent. Thus this kind of analogy influence does not persist for very long. This is also a feature of some analogical inductive logics considered in the next section.

\section{The Problem of Analogical Inductive Inference} \label{analogy}

Carnap's basic system of inductive logic can express analogical influences only to a limited degree \cite{Carnap1959,Niiniluoto1981,Romeijn2006}. There have been many attempts to extend Carnap's original system, and the literature on analogical inductive logic includes many valuable contributions \cite{Achinstein1963b,Carnap1980,Costantini1983,Festa1997,Hesse1964,Hill2013,Kuipers1984,Maher2000,Maher2001,Maio1995,Niiniluoto1981,Romeijn2006,Skyrms1993,Spohn1981}. I am going to discuss some of those contributions in order to motivate my own.

The biggest obstacle to analogical inference in Carnap's system is Johnson's sufficientness postulate \eqref{johnson}. Johnson's sufficientness postulate makes it impossible that counts $n_k$ of outcomes $k$ other than $i$ influence the predictive probability of $i$. Skyrms \cite{Skyrms1993} suggests an extension of Carnapian inductive logic that keeps exchangeability but drops Johnson's sufficientness postulate. Skyrms' proposal is further studied and extended in \cite{Festa1997} and \cite{Hill2013}, and a similar model is developed for a different context (two families of predicates) in \cite{Romeijn2006}. The basic idea is to use mixtures of inductive methods \eqref{basic} in order to account for initial analogies between outcomes. This is equivalent to considering mixtures of Dirichlet distributions instead of Dirichlet distributions in the de Finetti representation. Skyrms discusses this idea in terms of a wheel of fortune, where observations of an outcome should also increase the predictive probability of nearby outcomes. Using an appropriate mixture of Dirichlet priors makes this possible. The resulting probability distributions are exchangeable but violate Johnson's sufficientness postulate.

The analogy influence exhibited by these kinds of inductive systems is transient. This is due to the fact that the corresponding prior probabilities are exchangeable. Exchangeability implies that the counts of one outcome can only have an indirect effect on the predictive probabilities of other outcomes. To see this, suppose that an outcome $k$ is followed by an outcome $i$. Then exchanging $i$ with some arbitrary outcome in the past does not affect the joint probability. Thus, the effect of counts of $k$ outcomes affects the probability of $i$ outcomes indirectly via the initial parameters in the mixture of Carnapian inductive logics. 

In order to get systems that exhibit a more permanent analogy influence, exchangeability has to be dropped in addition to Johnson's sufficientness postulate. The inductive systems of Costantini \cite{Costantini1983}, Kuipers \cite{Kuipers1984}, Niiniluoto \cite{Niiniluoto1981} and, to a certain extent, Spohn \cite{Spohn1981} develop inductive logics of this type. In these models, the predictive probabilities for outcome $i$ do not just contain the counts $n_i$ but may also have terms with counts $n_k$ of other outcomes. Each of these systems is interesting in its own right, but for none of them is it clear what the underlying symmetry assumptions are, or whether they exhibit interesting symmetries at all, and thus they seem a bit ad hoc. 

Another criticism of some of these inductive methods was put forward by Spohn \cite{Spohn1981} and is also expressed by Costantini \cite{Costantini1983}. Because counts of all outcomes may explicitly influence the predictive probabilities of an outcome $i$, the corresponding inductive logics generally violate a postulate known as `Reichenbach's axiom'. Reichenbach's axiom says that predictive probabilities have to converge to limiting relative frequencies of sample outcomes, provided that the limit exists. That is, if $X_1, X_2, \hdots$ is an infinite sequence of outcomes such that the limit $n_i/n$ exists as $n \to \infty$, then
$$
\lim_{n \to \infty}\mb P[X_{n+1} = i| X_1, \hdots, X_n] = p.
$$
Besides Spohn's own system, Carnap's basic system and Skyrms' analogical system meet Reichenbach's axiom.

I think this critique misses the point of certain forms of inductive inference. The inductive logics of Costantini and Niiniluoto may be appropriate when there are underlying probabilistic dependencies between the outcomes. If these dependencies are persistent, then Reichenbach's axiom should not hold. The dependencies will not be reflected in relative frequencies of outcomes, while predictive probabilities should make use of known dependencies. I discuss this point further in the context of our analogical inductive logic.

It is not known whether the inductive methods discussed so far can be derived from a set of axioms analogous to those underlying the Johnson-Carnap system. This is a significant gap in our knowledge. The set of axioms from which the Johnson-Carnap continuum of inductive methods \eqref{basic} is derivable completely specifies inductive assumptions at the observational level, making it easy to determine whether one's priors conform to them. None of the above models of inductive inference has been treated within this Carnapian paradigm. Maher's inductive logic is something of an exception \cite{Maher2000,Maher2001}. He presents a set of axioms for an inductive logic with two families of predicates. Maher himself discusses problems for the extension of the inductive logic to predicate families containing more than two predicates \cite{Maher2001}, so I confine my attention to the case of two predicates. I think that already in this case one point is in need of clarification. 

Here is a brief overview of Maher's proposal. Suppose we have two families with two predicates. In the language of random variables, this means that we have two sequences of random variables $V_1,V_2, \hdots$ and $W_1,W_2, \hdots$, where each random variable can take on two different values (the possible values being different for the $V$'s and the $W$'s). For instance, the first sequence might record whether the coin lands heads and tails, and the second sequence may state whether the coin is flipped with the right or the left hand. Maher then considers the so-called `$Q$-predicates' (`state descriptions' in Carnap's terminology). The $Q$-predicates are all possible combinations of basic predicates from the two families. Again in the language of random variables, this means that we consider the sequence of pairs $Z_n = (V_n,W_n)$. The random vector $Z_n$ takes on pairs of values. Since the random variables $V_n$ and $W_n$ are binary, $Z_n$ can take on four values.

Maher \cite{Maher2000} assumes that the infinite sequence $Z_1, Z_2, \hdots$ is exchangeable. It follows from this that its probability distribution has a de Finetti representation. Maher's basic idea can then be described as follows. The de Finetti representation implies that we can construct the probability distribution of $Z_1, Z_2, \hdots$ by putting a prior distribution over the set of possible chances. Since the $Z_n$ can take on four different values, the set of possible chances is the three-dimensional simplex $\Delta_4 = \{(x_1, \hdots, x_4) \in \mathbb{R}^4|x_1,\hdots,x_4 \geq 0, x_1 +\hdots + x_4 = 1 \}$. Following an idea by Carnap \cite{Carnap1959}, Maher considers the subset of probability distributions in $\Delta_4$ where the two families of predicates are probabilistically independent. This is the set of all $(x_1, \hdots, x_4) \in \Delta_4$ such that $x_1 = (x_1 + x_2)(x_1 + x_3)$, which defines a two-dimensional surface in $\Delta_4$ that is known as the `Wright manifold' in population genetics.\footnote{After the population geneticist Sewall Wright. The Wright manifold is the set of probabilities that make the alleles at different genetic loci independent.} 

If the prior on $\Delta_4$ is a Dirichlet distribution, as in Carnap's basic system, then any two-dimensional surface in $\Delta_4$ has probability zero, since the Dirichlet distribution is absolutely continuous with respect to Lebesgue measure on $\Delta_4$. Thus, the Wright manifold has probability zero. Now, Carnap and Maher propose to look at a mixture between a Dirichlet distribution and a distribution that puts full weight on the Wright manifold. The resulting inductive logic is a mixture of Carnap's basic system on the random variables $Z_n$ and the product of Carnap's basic systems on the random variables $V_n$ and $W_n$. The former terms correspond to the hypothesis that the two predicate families are dependent and the product of the latter two terms to the hypothesis that they are independent. Using the de Finetti representation, Maher also provides an axiomatic basis from which this inductive method can be derived. He also shows with the help of examples that the resulting system seems to lead to plausible numerical results that capture certain analogy influences.

What type of analogy influences is this model supposed to capture? Maher wants to say that some of the $Q$-predicates are more similar than others, namely those that share at least one underlying predicate from the two families. If we denote the four combinations of values by $Q_1=(0,0)$, $Q_2= (1,0)$, $Q_3=(0,1)$ and $Q_4 = (1,1)$ then $Q_1$ is similar to $Q_2$ and $Q_3$, $Q_2$ to $Q_1$ and $Q_4$, $Q_3$ is similar to $Q_1$ and $Q_4$, and $Q_4$ is similar to $Q_2$ and $Q_3$. Maher's goal is to have an inductive logic that respects the analogies based on these similarities. But it is difficult to see the reason why placing positive prior probability on the Wright manifold should achieve this. There is no straightforward relationship between considering the two predicate families as independent and the intended analogies.

The one reason I can see is the following. The similarity relationships between the $Q$-predicates described in the previous paragraph yield four edges in $\Delta_4$ between the vertices that are considered similar. These edges are part of the Wrigth manifold. If one wishes to reflect the analogies between the $Q$-predicates in one's prior, then one's prior distribution over $\Delta_4$ should, presumably, place a sufficient amount of probability weight close to the four edges. One way to achieve this is by distributing probabilities in an appropriate way on the Wright manifold. But this is neither necessary nor sufficient. We can endow the Wright manifold by assigning positive probability only to the barycenter of $\Delta_4$ (which is an element of the Wright manifold) and probability zero to all the other points in the Wright manifold. In this case, the overall prior over $\Delta_4$ may not place the required probability weight close to the four edges. On the other hand, we may do exactly that without having to assign positive probability to the Wright manifold. Thus, even though it may work in some cases, assigning positive probability to the Wright manifold does not seem to be a principled solution to the analogy problem, which would characterize priors over $\Delta_4$ that assign a sufficient probability weight to the four edges between analogous $Q$-predicates. 

\section{Extending Partial Exchangeability} \label{partial}

The brief discussion in the previous section should make it clear that there are many forms of analogical inference. Each  form of analogical inference merits study, and existing inductive logics vary in their degree of solving analogical inference problems successfully. In the remainder of this paper I would like to propose one form of analogical inductive inference that is based on de Finetti's ideas about partial exchangeability and that can be solved within the Carnapian paradigm.

Recall that partial exchangeability looks at situations with outcomes of different types. This inductive situation can be illustrated with an example that Achinstein used to criticize Carnap's original inductive logic \cite{Achinstein1963b}. In this example we observe whether or not different types of metal conduct electricity. We might, for instance, look at osmium, platinum and rhodium. These three metals are the types in de Finetti's setup. Each type may or may not conduct electricity. This defines two outcomes. The analogy between types comes from the fact that they share certain significant chemical properties. Because of the analogy between types, it is reasonable to think that instances where osmium and rhodium where observed to conduct electricity are relevant for predictions of whether platinum conducts electricity. In this case, de Finetti's theory of partial exchangeability may be applied with a prior that reflects these analogies.

Partial exchangeability has a similar effect on analogical inferences as exchangeability: analogy is transient and vanishes in the limit. This makes sense in the example of flipping two coins. The similarity between the two coins may influence one's early judgements, but if there are no underlying dependencies between the coins the influence of similarity judgements will diminish. This is reflected by the fact that Reichenbach's axiom holds for predictive probabilities. But what if there are persistent dependencies between types? This might arguably be the case in the example of whether different metals conduct electricity, since there presumably is an underlying common cause for the relevant outcome.
Another example  can be constructed by considering the success of medical trials among males and females. The types are male and female, and the outcomes (in the simplest case) are whether the trial was successful or not. Now, there might be an underlying chancy dependency between types that is influenced by environmental and other factors. 
If this dependency is permanent, this should be reflected in the analogical inductive logic. 

How might such an inductive logic look like? The basic setup has a sequence of outcomes $X_1, X_2, \hdots$ and a sequence of types $Y_1, Y_2, \hdots$. Suppose, for simplicity, that there are only two types. Predictive probabilities concern future outcomes and not future types. The predictive probability of observing outcome $i$ given that it is of type $1$ and given past observations may be given by
\begin{equation} \label{predSimple}
\mb P[X_{N+1} = i | {\bf X}_N, {\bf Y}_N, Y_{N+1} = 1] = \frac{n_{i1} + \beta n_{i2} + \alpha_{i1}}{N_1 + \beta N_2 + \sum_j \alpha_{j1}}.
\end{equation}
In this formula, ${\bf X}_n =(X_1, \hdots X_N), {\bf Y}_N = (Y_1, \hdots, Y_N)$ are the past observations of outcomes and types; $n_{ij}$ is the number of outcomes $i$ of types $j$; and $N_1$ and $N_2$ are the total number of observations of type $1$ and $2$. The $\alpha$ parameters have the same meaning as in Carnap's basic system \eqref{basic}. The $\beta$ parameter expresses the analogy influence of observations of type $2$ on observations of type $1$. If $\beta$ is positive, then $i$ observations of type $2$ will have a positive influence on the predictive probability. This indicates a judgement of positive analogy between types. Moreover, analogy is permanent---since $\beta$ is a constant, the analogy influence of type $1$ on type $2$ does not vanish as $n$ increases.

There are many ways in which the qualitative features of the predictive probability \eqref{predSimple} could be formalized. Is \eqref{predSimple} just a formula that exhibits some resemblance to Carnap's original system? Or is there some underlying rationale? To see what is going on, notice, in the first place, that de Finetti's notion of partial exchangeability will not in general allow predictive probabilities to be of the form as given in \eqref{predSimple}. Partial exchangeability implies the following. Suppose that $X_{N+1} = i, X_{N+2} = k, X_{N+3} = j$. The predictive probability of this sequence of outcomes, given the past and the sequence of types $Y_{N+1} = 1, Y_{N+2} = 2, Y_{N+3} = 1$, is equal to the predictive probability of the sequence $X_{N+1} = j, X_{N+2} = k, X_{N+3} = i$ (in order to get from the first sequence of outcomes to the second we only exchange two outcomes within the same type). Now suppose that $k = j$. Then the first sequence of outcomes is $X_{N+1} = i, X_{N+2} = j, X_{N+3} = j$ and the second is $X_{N+1} = j, X_{N+2} = j, X_{N+3} = i$. It is difficult to see how in this case counts of outcome $j$ of type $2$ can have a constant influence on the predictive probability of outcomes $j$ of type $1$. If it had, its effect would have to be balanced exactly against the joint probability for the second sequence, which may not work in general.\footnote{For a precise statement, see my \cite{Huttegger2015}, especially Corollary 2.} 

The same issue does not arise if $k \not=i,j$. Thus, it seems reasonable to weaken partial exchangeability in order to allow for persistent analogical influences. We let $p^n_{ikj,st} = \mb P[X_{N+1} = i, X_{N+2} = k, X_{N+3} = j | {\bf X}_n,{\bf Y}_n,Y_{N+1} = s, Y_{N+2} = t, Y_{N+3} = s]$. Then generalized partial exchangeability requires, in the first place, that
$$
p^n_{ikj,st} = p^n_{jki,st}
$$
whenever $k \not = i,j$ (if $k = i$ or $k = j$, equality may but need not hold). Furthermore, let $p^n_{ij,s} = \mb P[X_{N+1} = i, X_{N+2} = j | {\bf X}_n,{\bf Y}_n,Y_{N+1} = s, Y_{N+2} = s]$. Then generalized partial exchangeability requires, in the second place, that
$$
p^n_{ij,s} = p^n_{ji,s}
$$

The next section is devoted to showing that generalized partial exchangeability, together with some further assumptions, leads to an interesting analogical inductive logic.
 
\section{A New Analogical Inductive Logic} \label{logic}

The most important additional assumption that we need is a modification of Johnson's sufficientness postulate:
\begin{equation} \label{suff}
\mb P[X_{N+1} = i|{\bf X}_N, {\bf Y}_N, Y_{N+1} = j] = f_{ij}(n_{i1}, n_{i2},N_1, N_2)
\end{equation}For simplicity, we continue assuming that there are only two types (for a generalization to a finite number of types, see \cite{Huttegger2015}). The modified sufficientness postulate says that predictive probabilities for an outcome $i$ depend on $i$, its type, as well as on the observed counts of $i$ outcomes of both types. This is a natural way to allow for analogical influences between types.

We also need two technical postulates. The first one is a regularity assumption to the effect that all finite sequences of types and outcomes have positive probability; i.e., every finite pair of sequences $X_1,\hdots, X_N, Y_1, \hdots, Y_N$ has positive probability. Finally, we assume that future types do not give information about the outcome of the next trial. More specifically, 
\begin{align}  \label{suffAdd}
\mb P&[X_{N+1} = i|X_1, \hdots, X_N, Y_{N+1} = j]\\ &=\mb P[X_{N+1} = i|X_1, \hdots, X_N, Y_{N+1} = j, Y_{N+2} = k] \nonumber \\ &=\mb P[X_{N+1} = i|X_1, \hdots, X_N, Y_{N+1} = j, Y_{N+2} = k, Y_{N+3} = l]. \nonumber 
\end{align}
This condition is a significant restriction for the applicability of our inductive logic. For example, think of types as different medical treatments (as in a bandit problem) and of outcomes as success or failure. Then a success on the next trial might not be probabilistically independent of future treatments. 

Suppose now that $X_1, X_2, \hdots$ and $Y_1, Y_2, \hdots$ are two infinite sequences of outcomes and types for which the foregoing assumptions hold (generalized partial exchangeability, modified sufficientness postulate, regularity, and conditional independence \eqref{suffAdd}). Suppose, in addition, that trials within types are not independent, and that there are at least three outcomes.\footnote{Assuming independence has the same reason as in the case of the Johnson-Carnap continuum---independence means that there is no inductive learning. Since the sufficientness postulate is empty if there are only two outcomes, this case has to be treated separately, for example by assuming additivity of predictive probabilities. An alternative approach is proposed in \cite{Costantini1979}.} Then the following theorem is true:

\begin{theorem}
There exist positive constants $\alpha_{ij}$ and nonnegative constants $\beta,\gamma$ such that $N_1 + \beta N_2 + \sum_{i} \alpha_{i1} \not= 0, N_2 + \gamma N_1 + \sum_{i} \alpha_{i2} \not= 0$ and
\begin{align*}
\mb P[X_{N+1} = i| {\bf X}_N, {\bf Y}_N, Y_{N+1} = 1] &= \frac{n_{i1} + \beta n_{i2} + \alpha_{i1}}{N_1 + \beta N_2 + \sum_{i} \alpha_{i1}} \\
\mb P[X_{N+1} = i| {\bf X}_N, {\bf Y}_N, Y_{N+1} = 2] &= \frac{n_{i2} + \gamma n_{i1} + \alpha_{i2}}{N_2 + \gamma N_1 + \sum_{i} \alpha_{i2}}
\end{align*}
for all $N$ and all $0 \leq n_{ij} \leq N_j$.
\end{theorem}

This theorem follows from a more general result in my \cite{Huttegger2015} where I prove these assertions for more than two types and allow the total number of trials to be finite.

The sequence of predictive probabilities can be generated by an urn model (just like the predictive probabilities of Carnap's basic system are generated by a Polya urn). Since the predictive probabilities of our new inductive logic do not fix the probabilities of types, we may first choose a sequence of types at random from a distribution that assigns positive probability to each finite sequence of types. Assume that we also have an urn for each type containing balls labelled by the outcomes. The initial distribution of balls in urn $j$ depends on the prior parameters $\alpha_{ij}$. We now start choosing balls from urns following the sequence of types. Whenever we choose a ball from an urn, we put it back together with another label. If the urn is of type $1$, we put a ball with weight $\beta$ into the urn associated with type $2$. 

The most important difference between our new inductive logic and Carnap's basic system \eqref{basic} are the parameters $\beta,\gamma$. Are there any good reasons to think that they are analogy parameters? Let me mention two. First, it can be shown that $\beta$ is positive if 
$$
P[X_2 = i | X_1 = i, Y_1 = 2, Y_2 = 1] > \mb P[X_1 = i | Y_1 = 1].
$$
Furthermore, $\beta$ increases as $P[X_2 = i | X_1 = i, Y_1 = 2, Y_2 = 1]$ approaches $1$.\footnote{Similar relations hold for $\gamma$; see \cite{Huttegger2015}.} This means that we have analogy effects of type $2$ on type $1$ if observing an outcome of type $2$ makes it sufficiently more likely to observe the same outcome of type $1$. This is what one would expect of an analogical inference.

The second reason becomes relevant if there are more than two types. Consider the analogy parameters of two types with respect to a third one. If one parameter is larger than the other, then observing outcomes of the former type raises the probability of outcomes of the third type more than observing outcomes of the second type.\footnote{See Proposition 1 in \cite{Huttegger2015}.}

The inductive logic of Theorem 1 is open to various interpretations. If we interpret the parameters $\beta$ and $\gamma$ as analogy parameters, then it is plausible to require that $\beta,\gamma \leq 1$ since, arguably, every type is maximally analogous to itself. This idea can be captured by another postulate:
\begin{align*}
\mb P[X_2 = i &| X_1 = i, Y_1 = j, Y_2 = j] \\ &\geq \mb P[X_2 = i |  X_1 = i, Y_1 = k, Y_2 = j]
\end{align*}
This says that an observation of an outcome $i$ of type $j$ never has a lower effect on the predictive probability of that outcome when it is of type $j$ than observing an outcome $i$ of another type. It is easy to see that this forces the analogy parameters $\beta,\gamma$ to be between zero and one.

But we may also think of types in terms of different information sources that are used to predict probabilities of outcomes. In this case, $\beta$ and $\gamma$ express judgements about the reliability of the two sources. Consequently, if $\beta > 1$ the agent believes that the second information source is more trustworthy than the first one and that, accordingly, information from type $2$ observations should have more weight.

One feature of the inductive logic of Theorem 1 was already discussed earlier in a different context. Our new inductive logic violates Reichenbach's axiom whenever the analogy parameters $\beta$ and $\gamma$ are positive. In this case, predictive probabilities converge to a convex combination of relative frequencies of outcomes of the two different types. As remarked earlier, if the underlying process is not assumed to be essentially independent, this is what one should expect. Our inductive logic allows types to be probabilistically dependent throughout the process of observation, and so observations from other types don't necessarily cease to be relevant for predictive probabilities of one particular type. Thus, Reichenbach's axiom should not be postulated for this case.

\section{Conclusion}

One of the biggest advantages of our inductive logic is that there is a precise set of conditions from which it can be derived. These conditions can be thought of as the inductive assumptions that make the use of our analogical inductive logic adequate, provided that they are thought to be true. For most other analogical inductive logics the underlying assumptions are not as clear, which makes it difficult to apply them.

What I wish to emphasize is that there are different ways to reason analogically. Accordingly, there is going to be a variety of legitimate analogical inductive logics, and not just the one inductive logic that fully captures analogical reasoning. One basic distinguishing feature is suggested by the foregoing discussion. There are, on the one hand,  inductive logics where analogies reflect initial similarities but are washed out with increasing information. On the other hand, there are permanent analogical inferences such as in our inductive logic. Here, analogy persists with increasing information. Which type of analogy is appropriate depends on one's inductive assumptions.


\begin{thebibliography}{10}
\providecommand{\bibitemdeclare}[2]{}
\providecommand{\surnamestart}{}
\providecommand{\surnameend}{}
\providecommand{\urlprefix}{Available at }
\providecommand{\url}[1]{\texttt{#1}}
\providecommand{\href}[2]{\texttt{#2}}
\providecommand{\urlalt}[2]{\href{#1}{#2}}
\providecommand{\doi}[1]{doi:\urlalt{http://dx.doi.org/#1}{#1}}
\providecommand{\bibinfo}[2]{#2}

\bibitemdeclare{article}{Achinstein1963b}
\bibitem{Achinstein1963b}
\bibinfo{author}{P.~\surnamestart Achinstein\surnameend}
  (\bibinfo{year}{1963}): \emph{\bibinfo{title}{Variety and Analogy in
  Confirmation Theory}}.
\newblock {\sl \bibinfo{journal}{Philosophy of Science}} \bibinfo{volume}{30},
  pp. \bibinfo{pages}{207--221}. \doi{10.1086/287935}

\bibitemdeclare{book}{Carnap1950}
\bibitem{Carnap1950}
\bibinfo{author}{R.~\surnamestart Carnap\surnameend} (\bibinfo{year}{1950}):
  \emph{\bibinfo{title}{Logical {Foundations of Probability}}}.
\newblock \bibinfo{publisher}{University of Chicago Press},
  \bibinfo{address}{Chicago}. \doi{10.1007/978-3-7091-3142-8}

\bibitemdeclare{book}{Carnap1952}
\bibitem{Carnap1952}
\bibinfo{author}{R.~\surnamestart Carnap\surnameend} (\bibinfo{year}{1952}):
  \emph{\bibinfo{title}{The Continuum of Inductive Methods}}.
\newblock \bibinfo{publisher}{University of Chicago Press},
  \bibinfo{address}{Chicago}.

\bibitemdeclare{incollection}{Carnap1971b}
\bibitem{Carnap1971b}
\bibinfo{author}{R.~\surnamestart Carnap\surnameend} (\bibinfo{year}{1971}):
  \emph{\bibinfo{title}{A Basic System of Inductive Logic, Part 1}}.
\newblock In \bibinfo{editor}{\surnamestart {Rudolf Carnap}\surnameend} \&
  \bibinfo{editor}{\surnamestart {Richard C. Jeffrey}\surnameend}, editors:
  {\sl \bibinfo{booktitle}{Studies in Inductive Logic and Probability I}},
  \bibinfo{publisher}{University of California Press}, \bibinfo{address}{Los
  Angeles}, pp. \bibinfo{pages}{33--165}.

\bibitemdeclare{incollection}{Carnap1980}
\bibitem{Carnap1980}
\bibinfo{author}{R.~\surnamestart Carnap\surnameend} (\bibinfo{year}{1980}):
  \emph{\bibinfo{title}{A Basic System of Inductive Logic, Part 2}}.
\newblock In \bibinfo{editor}{Richard~C. \surnamestart Jeffrey\surnameend},
  editor: {\sl \bibinfo{booktitle}{Studies in Inductive Logic and Probability
  II}}, \bibinfo{publisher}{University of California Press},
  \bibinfo{address}{Los Angeles}, pp. \bibinfo{pages}{7--155}.

\bibitemdeclare{book}{Carnap1959}
\bibitem{Carnap1959}
\bibinfo{author}{R.~\surnamestart Carnap\surnameend} \&
  \bibinfo{author}{W.~\surnamestart Stegm\"uller\surnameend}
  (\bibinfo{year}{1959}): \emph{\bibinfo{title}{Induktive Logik und
  Wahrscheinlichkeit}}.
\newblock \bibinfo{publisher}{Springer}, \bibinfo{address}{Wien}. \doi{10.1007/978-3-7091-3142-8}

\bibitemdeclare{article}{Costantini1979}
\bibitem{Costantini1979}
\bibinfo{author}{D.~\surnamestart Costantini\surnameend}
  (\bibinfo{year}{1979}): \emph{\bibinfo{title}{The Relevance Quotient}}.
\newblock {\sl \bibinfo{journal}{Erkenntnis}} \bibinfo{volume}{14}, pp.
  \bibinfo{pages}{149--157}. \doi{10.1007/BF00166497}

\bibitemdeclare{article}{Costantini1983}
\bibitem{Costantini1983}
\bibinfo{author}{D.~\surnamestart Costantini\surnameend}
  (\bibinfo{year}{1983}): \emph{\bibinfo{title}{Analogy by Similarity}}.
\newblock {\sl \bibinfo{journal}{Erkenntnis}} \bibinfo{volume}{20}, pp.
  \bibinfo{pages}{103--114}. \doi{10.1007/BF00166497}

\bibitemdeclare{incollection}{Diaconis1980}
\bibitem{Diaconis1980}
\bibinfo{author}{P.~\surnamestart Diaconis\surnameend} \&
  \bibinfo{author}{D.~\surnamestart Freedman\surnameend}
  (\bibinfo{year}{1980}): \emph{\bibinfo{title}{De {Finetti's} Generalizations
  of Exchangeability}}.
\newblock In \bibinfo{editor}{Richard~C. \surnamestart Jeffrey\surnameend},
  editor: {\sl \bibinfo{booktitle}{Studies in Inductive Logic and Probability
  II}}, \bibinfo{publisher}{University of California Press},
  \bibinfo{address}{Los Angeles}, pp. \bibinfo{pages}{233--249}. 

\bibitemdeclare{article}{Diaconis1980a}
\bibitem{Diaconis1980a}
\bibinfo{author}{P.~\surnamestart Diaconis\surnameend} \&
  \bibinfo{author}{D.~\surnamestart Freedman\surnameend}
  (\bibinfo{year}{1980}): \emph{\bibinfo{title}{De {Finetti's} Theorem for
  {Markov} Chains}}.
\newblock {\sl \bibinfo{journal}{Annals of Probability}} \bibinfo{volume}{8},
  pp. \bibinfo{pages}{115--130}. \doi{10.1214/aop/1176994828}

\bibitemdeclare{article}{Festa1997}
\bibitem{Festa1997}
\bibinfo{author}{R.~\surnamestart Festa\surnameend} (\bibinfo{year}{1997}):
  \emph{\bibinfo{title}{Analogy and Exchangeability in Predictive Inferences}}.
\newblock {\sl \bibinfo{journal}{Erkenntnis}} \bibinfo{volume}{45}, pp.
  \bibinfo{pages}{89--112}. \doi{10.1007/978-94-011-5712-4-6}

\bibitemdeclare{article}{Finetti1937}
\bibitem{Finetti1937}
\bibinfo{author}{B.~\surnamestart de~Finetti\surnameend}
  (\bibinfo{year}{1937}): \emph{\bibinfo{title}{La prevision: ses lois logiques
  ses sources subjectives}}.
\newblock {\sl \bibinfo{journal}{Annales d l'institut Henri Poincar\'e}}
  \bibinfo{volume}{7}, pp. \bibinfo{pages}{1--68}.
\newblock \bibinfo{note}{Translated in Kyburg, H. E. and Smokler, H. E.,
  editors, {\it Studies in Subjective Probability}, pages 93--158, Wiley, New
  York, 1964}.

\bibitemdeclare{incollection}{Finetti1938}
\bibitem{Finetti1938}
\bibinfo{author}{B.~\surnamestart de~Finetti\surnameend}
  (\bibinfo{year}{1938}): \emph{\bibinfo{title}{Sur la condition d'equivalence
  partielle}}.
\newblock In: {\sl \bibinfo{booktitle}{Actualit\'es Scientifiques et
  Industrielles No. 739: Colloques consacr\'e \`a la th\'eorie des
  probabilit\'es, VIi\`eme partie}}, \bibinfo{address}{Paris}, pp.
  \bibinfo{pages}{5--18}.
\newblock \bibinfo{note}{Translated in Jeffrey, R. C., editor, {\it Studies in
  Inductive Logic and Probability II}, pages 193--205, University of California
  Press, Los Angeles, 1980}.

\bibitemdeclare{incollection}{Finetti1959}
\bibitem{Finetti1959}
\bibinfo{author}{B.~\surnamestart de~Finetti\surnameend}
  (\bibinfo{year}{1959}): \emph{\bibinfo{title}{La probabilita e la statistica
  nei raporti con l'induzione, secondo i dwersi punti di vista}}.
\newblock In: {\sl \bibinfo{booktitle}{Corso C.I.M.E su Induzione e
  Statistica}}, \bibinfo{publisher}{Cremones}, \bibinfo{address}{Rome}.
\newblock \bibinfo{note}{Translated in de Finetti, B, {\it Probability,
  Induction and Statistics}, chapter 9, Wiley, New York, 1974}.

\bibitemdeclare{article}{Fortini2002}
\bibitem{Fortini2002}
\bibinfo{author}{S.~\surnamestart Fortini\surnameend}, \bibinfo{author}{Lucia
  \surnamestart Ladelli\surnameend}, \bibinfo{author}{Giovanni \surnamestart
  Petris\surnameend} \& \bibinfo{author}{E.~\surnamestart Regazzini\surnameend}
  (\bibinfo{year}{2002}): \emph{\bibinfo{title}{On Mixtures of Distributions of
  {Markov} Chains}}.
\newblock {\sl \bibinfo{journal}{Stochastic Processes and their Applications}}
  \bibinfo{volume}{100}, pp. \bibinfo{pages}{147--165}. \doi{10.1016/S0304-4149(02)00093-5}

\bibitemdeclare{article}{Freedman1962}
\bibitem{Freedman1962}
\bibinfo{author}{D.~\surnamestart Freedman\surnameend} (\bibinfo{year}{1962}):
  \emph{\bibinfo{title}{Mixtures of {Markov} Processes}}.
\newblock {\sl \bibinfo{journal}{Annals of Mathematical Statistics}}
  \bibinfo{volume}{33}, pp. \bibinfo{pages}{114--118}. \doi{10.1214/aoms/1177704716}

\bibitemdeclare{article}{Hesse1964}
\bibitem{Hesse1964}
\bibinfo{author}{M.~\surnamestart Hesse\surnameend} (\bibinfo{year}{1964}):
  \emph{\bibinfo{title}{Analogy and Confirmation Theory}}.
\newblock {\sl \bibinfo{journal}{Philosophy of Science}} \bibinfo{volume}{31},
  pp. \bibinfo{pages}{319--324}. \doi{10.1086/288017}

\bibitemdeclare{article}{Hill2013}
\bibitem{Hill2013}
\bibinfo{author}{A.~\surnamestart Hill\surnameend} \&
  \bibinfo{author}{J.~\surnamestart Paris\surnameend} (\bibinfo{year}{2013}):
  \emph{\bibinfo{title}{An Analogy Principle in Inductive Logic}}.
\newblock {\sl \bibinfo{journal}{Annals of Pure and Applied Logic}}
  \bibinfo{volume}{64}, pp. \bibinfo{pages}{1293--1321}. \doi{10.1016/j.apal.2013.06.013}

\bibitemdeclare{unpublished}{Huttegger2015}
\bibitem{Huttegger2015}
\bibinfo{author}{S.~M. \surnamestart Huttegger\surnameend}
  (\bibinfo{year}{2015}): \emph{\bibinfo{title}{Analogical Predictive
  Probabilities}}.
\newblock \bibinfo{note}{Manuscript, University of California at Irvine}.

\bibitemdeclare{book}{Johnson1924}
\bibitem{Johnson1924}
\bibinfo{author}{W.~E. \surnamestart Johnson\surnameend}
  (\bibinfo{year}{1924}): \emph{\bibinfo{title}{Logic, Part III: The Logical
  Foundations of Science}}.
\newblock \bibinfo{publisher}{Cambridge University Press},
  \bibinfo{address}{Cambridge, UK}. \doi{10.1093/mind/XLI.164.409}

\bibitemdeclare{article}{Johnson1932}
\bibitem{Johnson1932}
\bibinfo{author}{W.~E. \surnamestart Johnson\surnameend}
  (\bibinfo{year}{1932}): \emph{\bibinfo{title}{Probability: The Deductive and
  Inductive Problems}}.
\newblock {\sl \bibinfo{journal}{Mind}} \bibinfo{volume}{41}, pp.
  \bibinfo{pages}{409--423}.

\bibitemdeclare{book}{Kuipers1978}
\bibitem{Kuipers1978}
\bibinfo{author}{T.~A.~F. \surnamestart Kuipers\surnameend}
  (\bibinfo{year}{1978}): \emph{\bibinfo{title}{Studies in Inductive
  Probability and Rational Expectation}}.
\newblock \bibinfo{publisher}{D. Reidel}, \bibinfo{address}{Dordrecht}. \doi{10.1007/978-94-009-9830-8}

\bibitemdeclare{article}{Kuipers1984}
\bibitem{Kuipers1984}
\bibinfo{author}{T.~A.~F. \surnamestart Kuipers\surnameend} (\bibinfo{year}{1984}):
  \emph{\bibinfo{title}{Two Types of Inductive Analogy by Similarity}}.
\newblock {\sl \bibinfo{journal}{Erkenntnis}} \bibinfo{volume}{21}, pp.
  \bibinfo{pages}{63--87}. \doi{10.1007/BF00176183}

\bibitemdeclare{incollection}{Kuipers1988}
\bibitem{Kuipers1988}
\bibinfo{author}{T.~A.~F. \surnamestart Kuipers\surnameend}
  (\bibinfo{year}{1988}): \emph{\bibinfo{title}{Inductive Analogy by Similarity
  and Proximity}}.
\newblock In \bibinfo{editor}{D.~H. \surnamestart Helman\surnameend}, editor:
  {\sl \bibinfo{booktitle}{Analogical Reasoning}}, \bibinfo{publisher}{Kluwer},
  \bibinfo{address}{Dordrecht}. \doi{10.1007/978-94-015-7811-0-14}

\bibitemdeclare{article}{Maher2000}
\bibitem{Maher2000}
\bibinfo{author}{P.~\surnamestart Maher\surnameend} (\bibinfo{year}{2000}):
  \emph{\bibinfo{title}{Probabilities for Two Properties}}.
\newblock {\sl \bibinfo{journal}{Erkenntnis}} \bibinfo{volume}{52}, pp.
  \bibinfo{pages}{63--91}. \doi{10.1023/A:1005557828204}

\bibitemdeclare{article}{Maher2001}
\bibitem{Maher2001}
\bibinfo{author}{P.~\surnamestart Maher\surnameend} (\bibinfo{year}{2001}):
  \emph{\bibinfo{title}{Probabilities for Multiple Properties: The Models of
  {Hesse} And {Carnap} And {Kemeny}}}.
\newblock {\sl \bibinfo{journal}{Erkenntnis}} \bibinfo{volume}{55}, pp.
  \bibinfo{pages}{183--216}. \doi{10.1023/A:1012952802676}

\bibitemdeclare{article}{Maio1995}
\bibitem{Maio1995}
\bibinfo{author}{M.~C. \surnamestart di~Maio\surnameend}
  (\bibinfo{year}{1995}): \emph{\bibinfo{title}{Predictive Probability and
  Analogy by Similarity in Inductive Logic}}.
\newblock {\sl \bibinfo{journal}{Erkenntnis}} \bibinfo{volume}{43}, pp.
  \bibinfo{pages}{369--394}. \doi{10.1007/BF01135379}

\bibitemdeclare{article}{Niiniluoto1981}
\bibitem{Niiniluoto1981}
\bibinfo{author}{I.~\surnamestart Niiniluoto\surnameend}
  (\bibinfo{year}{1981}): \emph{\bibinfo{title}{Analogy and Inductive Logic}}.
\newblock {\sl \bibinfo{journal}{Erkenntnis}} \bibinfo{volume}{16}, pp.
  \bibinfo{pages}{1--34}. \doi{10.1007/BF00219640}

\bibitemdeclare{article}{Romeijn2006}
\bibitem{Romeijn2006}
\bibinfo{author}{J.-W. \surnamestart Romeijn\surnameend}
  (\bibinfo{year}{2006}): \emph{\bibinfo{title}{Analogical Predictions for
  Explicit Similarity}}.
\newblock {\sl \bibinfo{journal}{Erkenntnis}} \bibinfo{volume}{2006}, pp.
  \bibinfo{pages}{253--280}. \doi{10.1007/s10670-005-0232-8}

\bibitemdeclare{article}{Skyrms1991}
\bibitem{Skyrms1991}
\bibinfo{author}{B.~\surnamestart Skyrms\surnameend} (\bibinfo{year}{1991}):
  \emph{\bibinfo{title}{Inductive Logic for {Markov} Chains}}.
\newblock {\sl \bibinfo{journal}{Erkenntnis}} \bibinfo{volume}{35}, pp.
  \bibinfo{pages}{439--460}. \doi{10.1007/BF00388296}

\bibitemdeclare{incollection}{Skyrms1993}
\bibitem{Skyrms1993}
\bibinfo{author}{B.~\surnamestart Skyrms\surnameend} (\bibinfo{year}{1993}):
  \emph{\bibinfo{title}{Analogy by Similarity in Hyper-{Carnapian} Inductive
  Logic}}.
\newblock In \bibinfo{editor}{J.~\surnamestart Earman\surnameend},
  \bibinfo{editor}{A.~I. \surnamestart Janis\surnameend},
  \bibinfo{editor}{G.~\surnamestart Massey\surnameend} \&
  \bibinfo{editor}{N.~\surnamestart Rescher\surnameend}, editors: {\sl
  \bibinfo{booktitle}{Philosophical Problems of the Internal and External
  Worlds}}, \bibinfo{publisher}{University of Pittsburgh Press},
  \bibinfo{address}{Pittsburgh}, pp. \bibinfo{pages}{273--283}.

\bibitemdeclare{article}{Spohn1981}
\bibitem{Spohn1981}
\bibinfo{author}{W.~\surnamestart Spohn\surnameend} (\bibinfo{year}{1981}):
  \emph{\bibinfo{title}{Analogy and Inductive Logic: A Note on {N}iiniluoto}}.
\newblock {\sl \bibinfo{journal}{Erkenntnis}} \bibinfo{volume}{16}, pp.
  \bibinfo{pages}{35--52}. \doi{10.1007/BF00219641}

\bibitemdeclare{article}{Zabell1982}
\bibitem{Zabell1982}
\bibinfo{author}{S.~L. \surnamestart Zabell\surnameend} (\bibinfo{year}{1982}):
  \emph{\bibinfo{title}{{W. E. Johnson's }``Sufficientness" Postulate}}.
\newblock {\sl \bibinfo{journal}{The Annals of Statistics}}
  \bibinfo{volume}{10}, pp. \bibinfo{pages}{1091--1099}. \doi{10.1214/aos/1176345975}

\bibitemdeclare{article}{Zabell1989}
\bibitem{Zabell1989}
\bibinfo{author}{S.~L. \surnamestart Zabell\surnameend} (\bibinfo{year}{1989}):
  \emph{\bibinfo{title}{The {Rule of Succession}}}.
\newblock {\sl \bibinfo{journal}{Erkenntnis}} \bibinfo{volume}{31}, pp.
  \bibinfo{pages}{283--321}. \doi{10.1007/BF01236567}

\bibitemdeclare{article}{Zabell1995}
\bibitem{Zabell1995}
\bibinfo{author}{S.~L. \surnamestart Zabell\surnameend} (\bibinfo{year}{1995}):
  \emph{\bibinfo{title}{Characterizing {Markov} Exchangeable Sequences}}.
\newblock {\sl \bibinfo{journal}{Journal of Theoretical Probability}}
  \bibinfo{volume}{8}, pp. \bibinfo{pages}{175--178}.  \doi{10.1007/BF02213460}

\bibitemdeclare{incollection}{Zabell2011}
\bibitem{Zabell2011}
\bibinfo{author}{S.~L. \surnamestart Zabell\surnameend} (\bibinfo{year}{2011}):
  \emph{\bibinfo{title}{Carnap and the Logic of Inductive Inference}}.
\newblock In: {\sl \bibinfo{booktitle}{Handbook of the History of Logic}},
  \bibinfo{publisher}{Elsevier}, \bibinfo{address}{Amsterdam}, pp.
  \bibinfo{pages}{265--309}.

\end{thebibliography}
\end{document}